# Terahertz and Optical Acceleration Techniques

*Franz X. Kärtner*
Center for Free-Electron Laser Science – CFEL,
Deutsches Elektronen-Synchrotron – DESY, Hamburg, Germany

**Abstract**
The use of terahertz (THz) and optical radiation for electron acceleration and manipulation of electron bunches has progressed over the last decade to a level where practical devices for THz guns, THz and optical acceleration modules and a wide range of beam manipulations have become possible. Here, we discuss recent progress in optical driven Terahertz generation and its use in charged particle acceleration and beam manipulation devices. The advantages of using shorter wavelength radiation for acceleration are in overcoming breakdown phenomena, therefore enabling higher acceleration gradients than in conventional RF-accelerators albeit with lower bunch charge. The lower pulse energies needed to power the smaller cross section of the accelerating structures is also advantageous. In addition, the shorter wavelengths enable tighter timing control of the generated electron bunches but in return also need more precise timing when multiple stage interactions are required. Early results on THz guns, beam manipulation devices and accelerator structures are discussed as well as basic working principles of dielectric laser accelerators.

## 1    Introduction

At RF frequencies where conventional sources (klystrons, etc.) are efficient, surface electric field gradients in accelerating structures are limited by RF induced breakdown. Empirically, the breakdown threshold has been found to scale as $E_s \sim f^{1/2} \tau^{-1/4}$ [1,2] where $E_s$ is the surface electric field, $f$ is the RF-frequency of operation and $\tau$ is the RF-pulse length indicating that higher frequencies and shorter pulse durations are clearly beneficial, which has been experimentally verified in many studies [3,4]. Additionally, low frequencies (i.e. GHz) inherently require long RF-pulses because a single RF-cycle is long (on the order of ns) and traditional sources work most efficiently when operating over a very narrow frequency spectrum (i.e. long pulse length). Practically, this results in a significant amount of average power coupled into the structure if a high repetition rate is used. Therefore, accelerator scientists explore increasingly higher frequency electromagnetic waves ranging from the low THz wavelength range [5]. all the way up to infrared wavelength range for dielectric accelerators [6,7,8]. While the generation of IR-radiation around 2 μm can be achieved today with very much standard laser technology at high efficiency, single-cycle and multi-cycle THz generation has only been developed to high efficiency over the last decade and is still limited in efficiency, especially for short high energy pulses, i.e. high peak power pulses. Therefore, this paper is organized as follows. Chapter 2 will briefly review the status of single and multi-cycle THz pulse generation with greater than MW peak power. In Chapter 3, we will discuss two principal structures for THz accelerators, the Segmented Terahertz Electron Accelerator and Beam Manipulation device called STEAM [9] and the well-known dielectrically loaded waveguide accelerator (DLW) [10]. STEAM devices can also efficiently accelerate electrons from rest and, therefore, can also serve as electron guns. Chapter 4 will discuss first results achieved in ultrafast electron diffraction using a DC-gun and a DLW as a bunching device. Chapter 5 will briefly review recent results with dielectric accelerating structures (DLA) driven by IR-radiation.

## 2 THz generation

To achieve high acceleration gradients, where very high frequency accelerators have an advantage compared to RF-accelerators, short and high-power pulses are necessary. For this task optical generation of THz radiation by optical rectification has a great advantage, since optical pulses can have very high peak power and short duration. Therefore, even if the conversion efficiency is as low as 1%, very high peak power and short duration THz pulses can still be generated. Albeit in the long run, one has to improve on efficiency. It turns out that the most appropriate, i.e. highest efficiency, optical rectification can be achieved in Lithium Niobate ($LiNbO_3$). Due to the scaling of the THz generation efficiency with frequency and THz-absorption, there is an optimum frequency for THz generation and frequencies of 0.2 to 1 THz seem to work best.

### 2.1 Single-cycle THz generation

For single-cycle THz generation, broadband phase matching is accomplished via the tilted pulse-front method [11] and a systematic study of this process has been performed in [12]. However, development of practical devices requires THz sources that reliably provide pulse energies in the µJ - mJ regime, which in turn require state-of-the-art pump laser systems and carefully designed optical transport lines. Kroh et al. thoroughly investigated both by experiments and simulations how spatio-temporal coupling of pump pulse parameters in tilted pulse-front based terahertz setups can be used to control the position of the "temporal focus", which is where the minimum pump pulse duration is reached inside the crystal [12]. This concept opens a pathway to pumping tilted-pulse-front setups with arbitrarily stretched pulses, which significantly simplifies transport lines for lasers with high peak intensity. This concept was experimentally demonstrated by efficiently pumping a tilted pulse-front THz source, see Fig. 1, with pulses stretched to 10 ps from a commercially available Yb-based amplifier system providing 410 fs pump pulses with up to 200 mJ energy centered at $\lambda = 1030$ nm and repetition rate of 52 Hz, see Ref. [13].

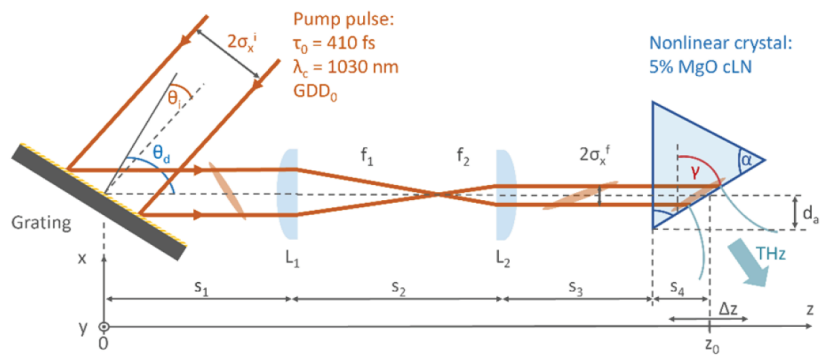

**Fig. 1:** Schematic drawing of the setup including crucial tuning parameters, after Ref. [13].

The THz setup was designed and optimized based on the considerations and procedures laid out in Ref. [12]. A congruent 5 % $MgO:LiNbO_3$ prism was cryogenically cooled to 82 K using liquid nitrogen. A reflective diffraction grating ($\rho = 1500$ l/mm, diffraction order m = 1) paired with a cylindrical telescope (M = 0.654 ± 0.002) was used to tilt the intensity front of the pump pulse and image the grating. The initial GDD on the pump pulse was controlled by tuning of the optical path between the diffraction gratings in the optical compressor of the laser system, while the crystal was positioned along the z-axis by use of a motorized translation stage. All THz pulse energies reported in this work refer to the usable energy measured outside of the dewar, corrected by the measured average transmission (≈ 54 %) through the black polyethylene cover (1.9 mm thickness) in front of the calibrated THz detector (THz 20, SLT Sensor- und Lasertechnik GmbH). THz pulses up to 0.4 mJ at a center frequency of about 0.3 THz have been extracted from the setup with typical beam profile and waveform shown in Fig. 2.

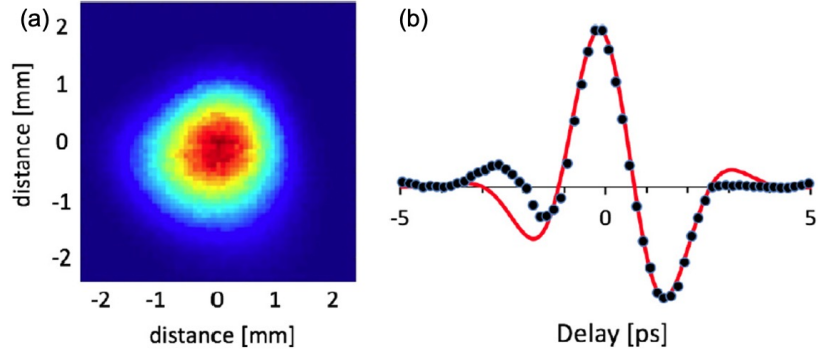

**Fig. 2:** (a) Measured THz beam profile at source. (b) THz waveform measured by electro-optic sampling.

## 2.2 Multi-cycle THz generation

As we will see in the next chapter, the DLW will need multi-cycle THz pulses preferably with multi-mJ pulse energy, 0.1-1 ns duration, and we also aim here for a center frequency of about 0.3 THz. For multi-cycle THz generation, quasi-phase matching using periodically-poled lithium niobate (PPLN) crystals is the preferred method since theory and numerical simulations have shown that it allows scaling to long interaction lengths with efficiencies reaching eventually multiple percent [14]. In early work, we reported generation of microjoule-level multi-cycle pulses, two orders of magnitude more energetic than previously reported, by cryogenic cooling of the PPLN and by using compressed pulses at 800 nm with a tailored bandwidth, resulting in efficiencies of about 0.1% [15]. This work, however, was limited both by the aperture of the available crystals (3 mm x 3 mm) and by the damage threshold of the material, which limits the incident intensity. In more recent work it was demonstrated that energy could be scaled up by another order of magnitude (to ~40 μJ) by using two pulses linearly chirped in time to reduce the intensity for a given fluence, with a delay chosen so that the instantaneous frequency difference between the two pulses satisfied the narrow quasi-phase matching condition of the PPLN [16]. Another order of magnitude (and record energy of 604 μJ) has very recently been achieved by tuning the phase structure of the chirped pulses and by increasing the aperture of the crystal to 10 mm x 15 mm in order to allow an increase in the incident energy at the same fluence [17]. These large aperture crystals, provided by Prof. Taira from the Laser Research Center at the Institute for Molecular Science in Japan, represent the state of the art in what is currently possible for PPLN. Figure 3 shows the spectral characterization of the multi-cycle THz pulses generated in collaboration with the LUX group at DESY on a Ti:Sapphire laser at 5 Hz with up to 1 J of available energy. While these results represent a significant step forward in optical generation of multi-cycle THz pulses, another similar step forward will be required to reach the 20 mJ of multi-cycle THz energy anticipated to be needed for a THz linear accelerator (LINAC). Research into this next step is on-going. However, it is expected that it will require, among other things, optimization of the spectral content and spatio-temporal profiles of the laser pulses [18], optimization of the output coupling of the THz from the crystals and recycling of the laser pulse energy in multiple crystals. Early results from simulations under development indicate that with these improvements, tens of millijoules of energy are possible from a 1 J laser pulse.

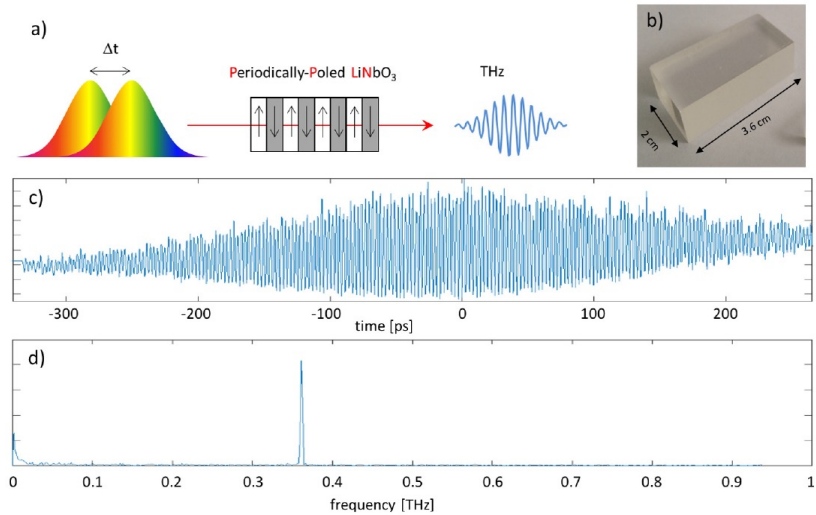

**Fig. 3:** (a) Chirp & delay concept (b) Large aperture PPLN donated by Prof. Taira from the Institute for Molecular Science in Japan. (c) Interferometric autocorrelation of the THz pulse. (d) Power spectrum of THz pulse obtained from Fourier Transform of (c), centered at 0.36 THz, Ref. [17].

## 3   THz guns, accelerators and beam manipulators

Over the last years, substantial THz acceleration of electrons from a DC-gun [5] or from rest [19] using relatively low energy THz pulses, in the few tens of μJ range, was demonstrated showing the feasibility of THz acceleration. Since then, quasi-monochromatic electron bunches with a few percent energy spread and 400 eV mean energy have been demonstrated from a parallel plate THz gun [20]. Based on these early results, we envisioned a general set of THz acceleration and beam modification devices shown in Fig. 4.

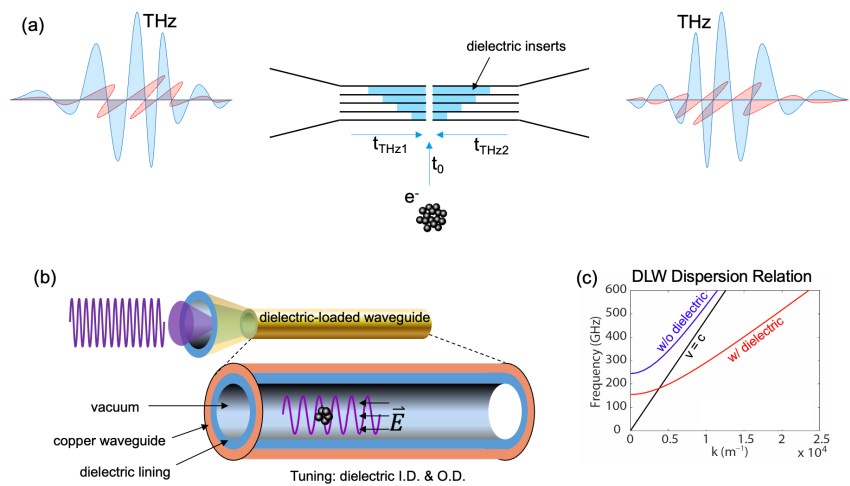

**Fig. 4:** (a) Principal setup of Segmented THz Accelerator and Manipulator – STEAM device powered by two transverse injected close to single-cycle THz pulses. (b) Principal setup of THZ-LINAC using a Dielectrically Loaded Waveguide (DLW). (c) Dispersion relation of DLW. The dielectric lining enables phase matching to a relativistic particle propagating on the light line v = c.

Figure 4(a) shows the general structure of a segmented THz accelerator and beam manipulator – STEAM, see Ref. [21]. The STEAM device is powered by two single-cycle THz pulses injected into a stack of THz waveguides, where the partial THz pulses in each waveguide are properly delayed such that at the time the electron bunch passes this waveguide section the electric or magnetic fields of the single-cycle pulses constructively interfere and or cancel. If the electric fields interfere constructively, the magnetic components cancel and we obtain an accelerator when the electron bunch passes at the maximum field or a compressor if the electron passes at a zero crossing. If the pulses are timed such that the magnetic components cancel and if the electron bunch passes at the maximum of the B-field, the device acts as a deflector. A device with many THz waveguides implemented can therefore overall execute a complex 6D-phase space manipulation. A STEAM device starting with a cathode and photo injector pulse can also function as a THz-gun accelerating electrons from rest to relativistic speeds, see Ref. [21]. Figure 4(b) shows a multi-cycle THz powered dielectrically loaded waveguide (DLW). The DLW enables phase matching to a relativistic and also sub-relativistic particles when properly dimensioned, i.e. proper choice of dielectric and inner (I.D.) and outer diameter (O.D.), see Fig. 4(c). Thus, after generating a relativistic electron bunch with a STEAM device, it can be injected into a THz-DLW using either a proper coupling structure or transforming the typically linearly polarized THz-pulses into a radially polarized beam using a segmented waveplate, see Ref. [5], which is the concept for a complete THz based electron and X-ray source described in Refs. [22,23].

### 3.1  Segmented THz Accelerator and Manipulator - STEAM

The segmented terahertz waveguide device acting as an electron accelerator and manipulator is capable of performing multiple high-field operations on the 6D-phase-space of ultrashort electron bunches. We have demonstrated this device powered by single-cycle, 280 GHz pulses and demonstrated > 30 keV acceleration, streaking with <10 fs resolution, focusing with >2 kT/m field strength, compression to ~100 fs as well as real-time switching between these modes of operation, see Ref. [9]. The yet rather basic device with only three segments was used to accelerate a 55-keV electron bunch from a DC-gun by more than 30 keV, see Fig. 5. This is the case when the two THz pulses entering the waveguide from both sides superimpose in the center of the device such that the electric fields pointing in forward direction superimpose constructively, whereas the magnetic field cancels. If one of the waves is shifted by half a wavelength, the electric field cancels and the magnetic fields add up constructively, leading to a deflecting Lorentz force on the electrons for temporal electron bunch characterization.

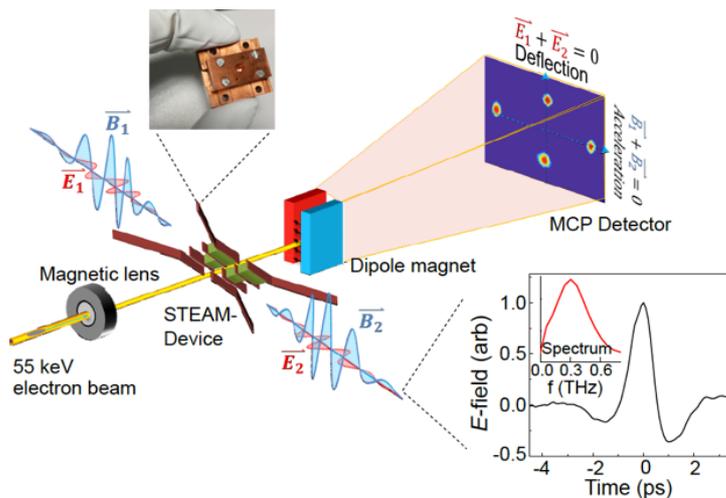

**Fig. 5:** Acceleration and manipulation of an electron bunch from a 55 keV DC-gun using a STEAM device with dipole magnet for energy measurement on an MCP. The device is driven by two single-cycle 0.3 THz pulses [9].

The electrons are injected In the electric mode with maximum acceleration field. The peak fields reached ~70 MV/m and produced more than 30 keV acceleration using the available ~2×6 μJ of coupled THz energy, see Fig. 6(a). In this configuration, the deflection was minimized due to the cancellation of the magnetic fields. At the zero-crossing of the electric field, the temporal field gradient is maximized resulting in a time-varying electric field and thus a velocity gradient that induces either compression or stretching of the electron bunch as it propagates, see Fig. 6(b). A minimum duration of ~100 fs FWHM was achieved. The temporal electric-field gradient at the zero crossing is intrinsically coupled to a transverse spatial gradient in the magnetic fields which results in focusing or defocusing of the electron beam, see Fig. 6(c). Peak focusing gradients reached over 2 kT/m, approaching those found in an active plasma lens.

In the magnetic mode, the relative timing of the THz fields is different from that of the electric mode by a half period resulting in reinforcement of the magnetic and cancellation of the electric fields. When electrons are on the crest of the magnetic field, the deflection is maximized and the acceleration (or deceleration) is minimized. Electrons arriving at the zero-crossing of the THz magnetic field experience a deflection that is a steep function of time, effectively streaking the bunch and projecting the temporal electron charge profile onto the spatial dimension of the MCP detector with sub-10 fs temporal resolution, see Fig. 6(d).

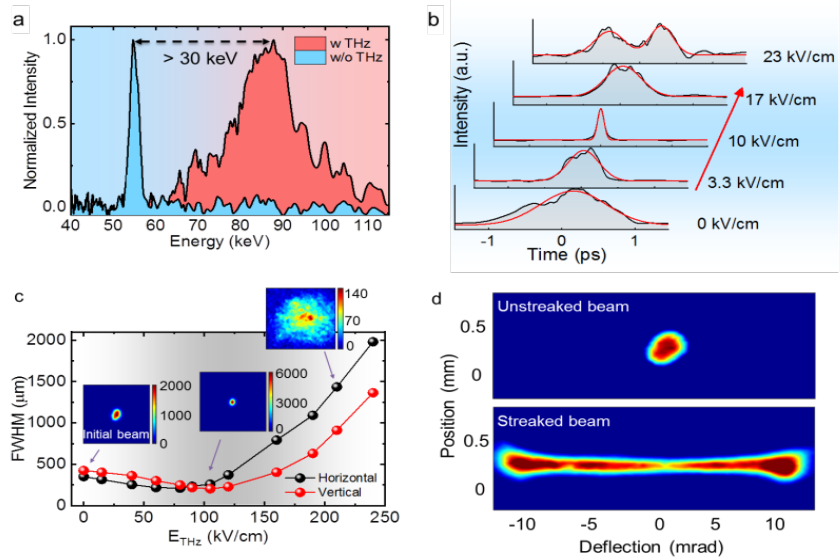

**Fig. 6:** Functions of the STEAM device. (a) Acceleration of electron beam with electron energy spectra for input beam (blue curve) and accelerated beam (red curve). (b) Temporal compression of the electron pulses as a function of the THz field in the rebunching mode. (c) Transverse electron beam focusing as a function of the THz field in the focusing mode. (d) Images of the electron beam on the detector with and without the THz deflection field in the streaking mode, Ref. [9].

## 3.2 THz LINAC using a DLW

Dielectric loaded waveguides (DLWs) driven by multi-cycle THz pulses hold great promise as compact linear accelerators (LINACs) due to their ability to sustain higher breakdown fields at THz frequencies compared to conventional RF components. Precise control of the THz pulse's phase and group velocities within the DLW can be achieved by adjusting the dimensions of the dielectric tube. Optimization of the DLW parameters has to take various factors into account like initial electron energy, THz pulse energy available, THz pulse width, DLW dimensions, etc. A comprehensive analytical and numerical guideline for cylindrical DLW LINACs aimed at maximizing the final kinetic energy of accelerated electrons can be found in Ref. [10]. Additionally, graphic representations are introduced to visualize optimal designs for varying initial electron and THz pulse energies. The provided guideline figures enable designers to tailor the accelerator to specific requirements, paving the way for potential advancements in THz-driven particle acceleration. Figure 7 shows the basic structure of a DLW. Critical dimensions are the vacuum diameter and the dielectric thickness.

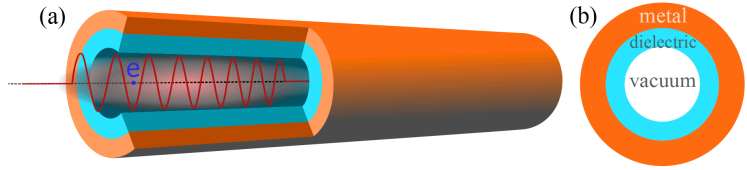

**Fig. 7:** (a) DLW structure comprising a dielectric tube (light blue) surrounded by a metallic layer (orange) functioning as a THz LINAC (b) Cross-sectional view of the DLW, Ref [10].

The dimensions can be chosen such, that the lowest order modes are the transverse electric mode $TE_{11}$ and transverse magnetic mode $TM_{01}$. By proper excitation of the DLW one can suppress the $TE_{11}$-mode and excite only the $TM_{01}$-mode and adjust its phase velocity close to the speed of light, such that relativistic particles are in phase with this mode. The electric field distribution of the $TM_{01}$–mode is shown in Fig. 8. The $TM_{01}$–mode shows a slow group velocity being adjusted close to its cutoff. This leads to field enhancement in the DLW when powered with long multi-cycle THz pulses, see Ref. [10].

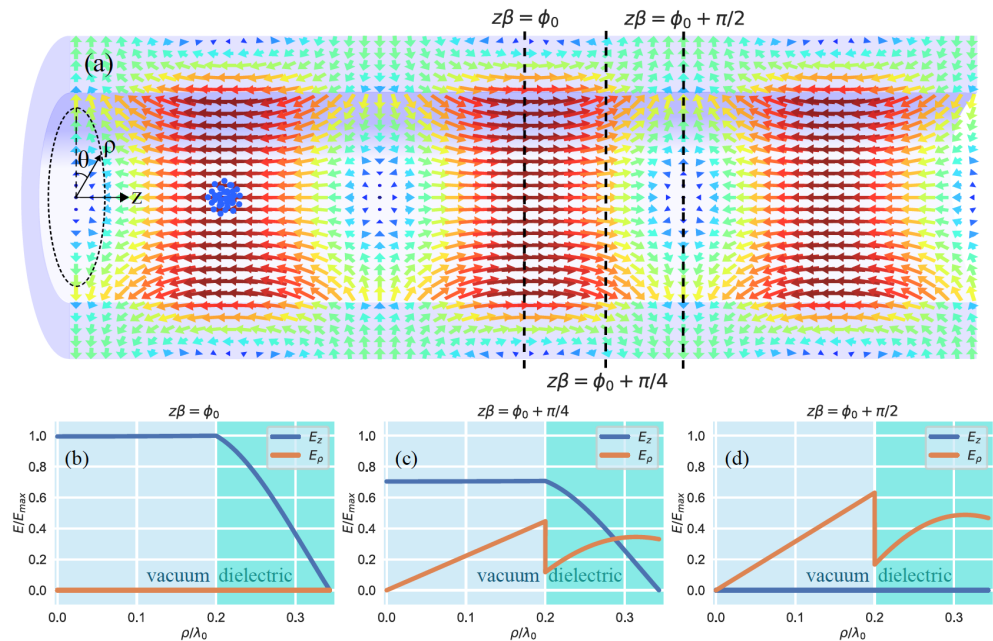

**Fig. 8:** (a) Electric field distribution of the TM01-mode within the DLW for a cross section along the waveguide at a fixed time t. The filed distribution is radially symmetric around the propagation axis. (b-d) show the fields over the waveguide cross-section at the different propagation phases indicated in Fig. (a). The dielectric thickness to vacuum radius ratio has been deliberately selected to align the phase velocity with the speed of light, Ref. [10].

# 4 THz-powered ultrafast electron diffractometer

Terahertz based electron manipulation has the lowest threshold in terms of THz-pulse energy needed, and, therefore holds great promise as a technology for manipulating and driving the next-generation of compact ultrafast electron sources [24]. In this chapter, we demonstrate an ultrafast electron diffractometer with THz-driven pulse compression using a DLW [25]. The electron bunches from a conventional DC gun are compressed by a factor of 10 and reach a duration of ~180 fs (FWHM) with 10,000 electrons/pulse at a 1 kHz repetition rate. The resulting ultrafast electron source is used in a proof-of-principle experiment to probe the photoinduced dynamics of single-crystal silicon. The THz-compressed electron beams produce high-quality diffraction patterns and enable observation of the ultrafast structural dynamics with improved time resolution. These results validate the maturity of THz-driven ultrafast electron sources for use in precision applications.

Figure 9 shows the experimental setup. The electron beam from a 53 keV photo-triggered DC gun is compressed by a multi-cycle THz powered DLW device. Its pulse duration is analyzed by a STEAM device not shown for simplicity.

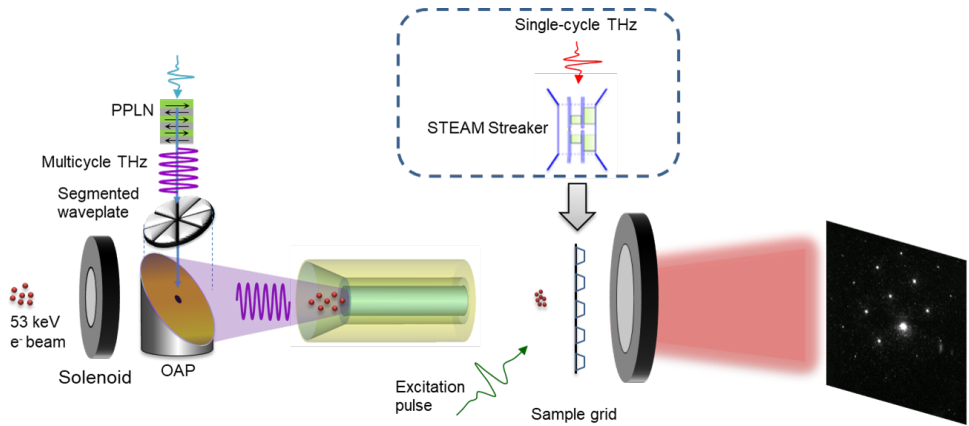

**Fig. 9: Experimental setup of a THz-powered electron diffractometer.** A small fraction of the 1030-nm infrared optical beam is converted to 257-nm based on two-stage second harmonic generation. The 257-nm UV pulse is directed onto a gold photocathode generating electron pulses, which are accelerated to 53 keV by the DC electric field with around 1 fC of charge. The same infrared laser also drives a multicycle THz generation stage, two single-cycle THz stages and pump laser for the DLW manipulator, the STEAM streaker and sample excitation, respectively. The STEAM streaker and the sample are on the same manipulator which can be exchanged for checking the pulse duration or the ultrafast electron diffraction (UED) experiment, Ref. [25].

Ultraviolet (UV) pulses for photoemission in the DC gun, multi-cycle THz pulses to drive the DLW device, single-cycle THz pulses to drive the STEAM device and optical pump laser pulses for the sample excitation are all created using a single, infrared Yb:KYW laser system producing 4-mJ, 650-fs, 1030-nm pulses at 1 kHz repetition rate. The UV pulses are generated by two successive stages of second harmonic generation (SHG); 50 ps long multi-cycle THz pulses are generated by intra-band difference frequency generation in a 5 mm long periodically poled lithium niobate (PPLN) crystal; single-cycle THz pulses are generated via the tilted-pulse-front method in a $LiNbO_3$ prism; and 515 nm pump pulses for the sample excitation are generated via SHG inside a BBO crystal. The linearly-polarized multi-cycle THz beam is converted to a radially-polarized beam, i.e. $TM_{01}$–mode, via a segmented waveplate with 8 segments. It is then coupled into the DLW device collinearly to the electron propagation using an off-axis-parabolic mirror and horn structure that concentrated the THz field into the DLW. The DLW design consists of a cylindrical copper waveguide of diameter 790 μm and a dielectric layer of alumina ($Al_2O_3$, THz refractive index n=3.25) with a wall thickness of 140 μm.

For electron compression, 0.5-mJ laser pulses are used for multi-cycle THz generation and the rest (3.5 mJ) is used for single-cycle THz generation and sample excitation. The single-cycle pulses centered at 300 GHz have an energy of around 2*3 µJ. They are injected into the STEAM streak camera [25] for electron pulse duration characterization.

The electrons are positioned at the negative zero crossing. Compression of the electron bunch is based on "velocity bunching" [26], where the electric field imparts a longitudinal, temporally-varying energy gain resulting in a velocity gradient that causes compression of the electron bunch as it propagates. Specifically, the electron bunch experiences acceleration at the tail and deceleration at the head, but no average energy gain. Due to the low phase velocity, the on-axis field is largely suppressed with most of the field present around the dielectric region of the waveguide. Performance of the buncher also improves with the energy of the injected electrons. The faster electron bunches match higher THz phase velocities that result in fields with a flatter distribution and a higher peak value in the center.

Compression of the bunch is shown in Fig. 10(a). At maximum compression the electron bunch duration was reduced by a factor of 10 to around 180 fs (FWHM) measured by the STEAM streaker. The STEAM streaker uses the magnetic field of the THz to induce the transverse deflection of the electron beam as discussed in the previous chapter. When the electrons sweep the zero-crossing of the field, they experience a strong deflection as a function of delay, enabling the measurement of the temporal bunch profile by mapping it onto the spatial dimension of a detector. The resultant temporal resolution is about 10 fs. Due to its compactness, the device could be directly mounted onto the UED sample manipulator ensuring that the pulse duration was measured at the sample position. Varying the energy, and hence the peak field, of the THz pulse allowed tuning of the longitudinal location of the temporal focus to coincide with the sample. As shown in simulations, Fig. 10(b), for lower fields, the temporal focus is located beyond the target position; while for stronger fields the bunch focused before the target. Analogously to optics, the tightness of the focusing determines the size of the focus. Since the THz pulse energies required to compress the bunch are much lower than what was available, much shorter bunch durations can be achieved simply by increasing the field strength, see Fig. 10(b). The mechanical design of the proof-of-principle setup, however, was not optimized for minimizing the electron bunch durations, and the long distance between buncher and sample limited the pulse duration achievable.

To determine the timing stability of the system, the spatial position of the electrons on the detector was monitored. Variation in the timing between the electrons and the bunching field induces a net acceleration or deceleration in the bunch which results in a delay relative to the streaking field. Jitter in the time of arrival of the streaking field would have a similar effect. These relative timing variations then manifest themselves as spatial deflections on the detector. Figure 10(c) shows the reconstructed relative timing jitter over the 5 minutes required for collection of one set of UED data. The RMS deviation was less than 5 fs even in the absence of any stabilization hardware. These timing fluctuations can be further reduced by stabilizing the laser beam pointing and thermal drifts of the system.

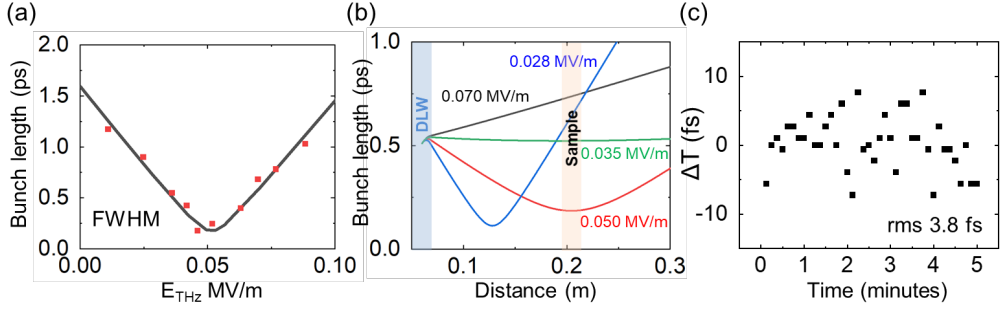

**Fig. 10:** Compression of electron bunch and system timing jitter. (a) Measured (red square) and simulated (black line) electron bunch length as a function of the applied THz field in the compression mode. (b) Simulated bunch length along the propagation direction with different longitudinal THz field strength. (c) Measured timing jitter between the zero crossing of the longitudinal THz electric field and the laser pulses. The measured timing jitter of about 3.8 fs RMS deviations reveals the excellent longer-term stability of the setup without any active stabilization, Ref. [25].

To demonstrate the performance of the setup, we measured the ultrafast (Debye-Waller) dynamics during heating of a 35 nm freestanding, single-crystalline silicon sample. Figure 11(a) shows the high-quality diffraction signal collected with 1 s exposure time. The slight distortion of the diffraction pattern visible is mainly caused by misalignment of the focusing solenoid which can be eliminated by upgrading the setup to provide motorization of the solenoid position. The sample is photoexcited with 515 nm laser pulses at a fluence of around 5 mJ/cm$^2$, well below the damage threshold. The recovered structural dynamics is shown in Fig. 11(b). The exponential fit of τ = 1±0.2 ps is slightly longer than previous measurements (0.88 ps) [27], which is mainly due to the pulse duration of the pump laser (~0.5 ps) that limits the overall system temporal resolution. The dynamics measured with the uncompressed electron beam shows a much longer decay time which is limited by the duration of the uncompressed electron bunch, see Fig. 10(a).

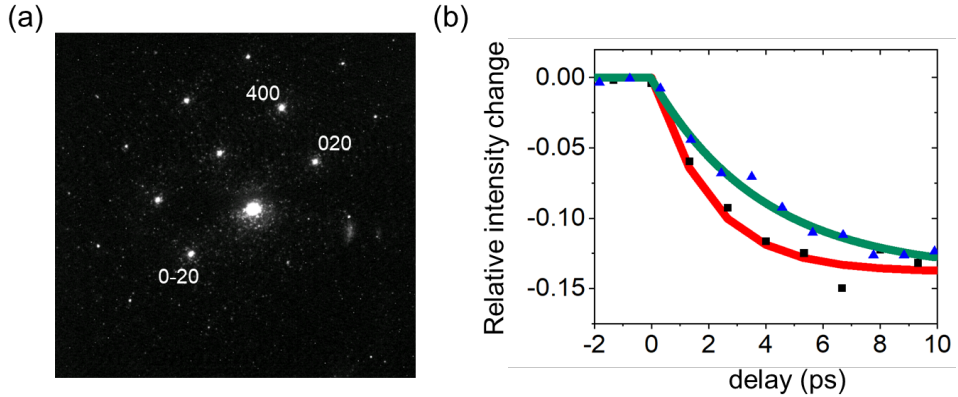

**Fig. 11:** Ultrafast electron diffraction on silicon. (a) Electron diffraction images of 35 nm single-crystalline silicon with a face-centered cubic structure. The data is collected by an MCP detector with 1 s exposure time. (b) The relative intensity changes of the 400 diffraction spots as a function of the time delay under the incident laser fluence of around 5 mJ/cm$^2$ with compressed electron bunch (black square) and uncompressed electron bunch (blue triangle). Each fit represents a function of single exponential decay with a time constant of 1±0.2 ps (red) and 2.6±0.3 ps (green), Ref. [25].

The results shown here can be improved upon in several ways. Most important are reducing the electron bunch duration, in order to improve the temporal resolution; and increasing the electron bunch energy, which is needed to enable study of samples which are thicker or are in a gas phase.

The first can be done by reducing the duration of the UV photoemission pulse, for example by implementing an optical parametric amplifier. The second can be done by implementing an additional THz-powered module for acceleration, as previously demonstrated with the STEAM device [9]. Adding an accelerator also improves the bunch duration. In this configuration, both modules can contribute to bunch compression.

## 5    Direct acceleration by optical lasers in dielectric structures

A detailed review of dielectric laser accelerators (DLAs) can be found in Ref. [8]. One of the basic DLA-structures are optical gratings, that can be lithographically fabricated on integrated optical chips, see Fig. 12.

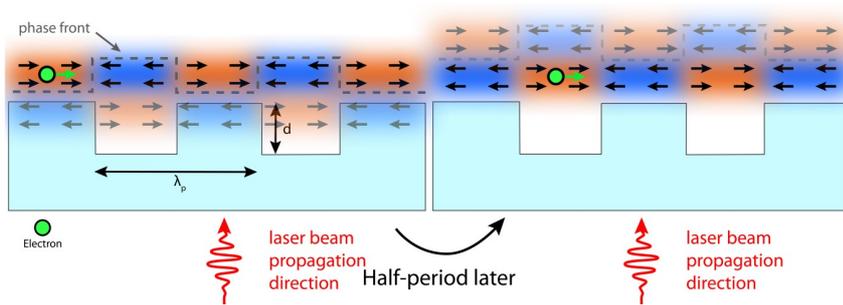

**Fig. 12:** Electron acceleration in the nearfield of an optical grating and schematic illustration of phase matching, courtesy of P. Hommelhoff.

Similar to the STEAM device the structure is illuminated sideways with optical radiation, typically 100 fs to picosecond pulses. To achieve phase matching the grating teeth spacing is adapted to the optical wavelength and initial speed of the electron, such that phase matching over longer propagation distances are achieved. First implementations of this scheme have been demonstrateed about a decade ago [6, 7]. Due to the high damage threshold of dielectrics at optical frequencies and low optical nonlinearities GV/m accelerating gradients are conceivable corresponding to acceleration gradients 1–2 orders of magnitude higher than conventional copper RF accelerators. At higher fields, i.e. intensities, optical nonlinearities in the gratings and eventual integrated optical waveguides become however excessive. Acceleration gradients as high as 850 MeV/m for relativistic DLAs and 370 MeV/m for subrelativistic DLAs have been demonstrated. Leveraging these high gradients over long interaction lengths to produce high energy gains at subrelativistic energies has been difficult due to confinement challenges. To mitigate this problem alternative phase focusing (APF)-DLAs [28] were introduced, see Fig. 13.

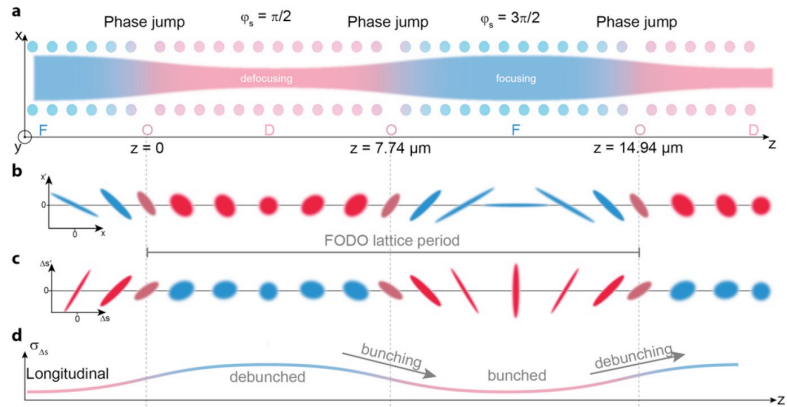

**Fig. 13:** Complex optical electron phase-space control in alternating phase focusing. (a) Sketch of the electron-beam envelope (coloured band) under the influence of the optical nearfield generated by the dielectric nanostructure. Filled circles indicate nanopillars extruded out of plane (y-direction). (a), (b), Evolution of a particle ensemble in transverse phase space relative to the structure layout: The electron distribution at z = 0 is compressed in transverse phase space (x extent decreasing), while in this D cell transversally defocusing forces already act on the electrons. (c), (d), The longitudinal forces, 90° out of phase relative to the transverse forces, act similarly on the longitudinal phase-space distribution, according to Ref. [29].

APF-DLAs are designed to act as alternating phase focusing lattices, where electrons, depending on the electron-laser interaction phase, will alternate between opposing longitudinal and transverse focusing and defocusing forces. By incorporating fractional period drift sections that alter the synchronous phase between $\pm 60°$ off crest, electrons captured in the designed acceleration bucket experience half the peak gradient as average gradient while also experiencing strong confinement forces that enable long interaction lengths. APF accelerators with interaction lengths up to 708 $\mu$m and energy gain up to 23.7 $\pm$ 1.07 keV FWHM, a 25% increase from starting energy, were demonstrated, showing the ability to achieve substantial energy gains with subrelativistic DLA, see Ref. [29].

## 6     Conclusion and Outlook

THz generation, THz acceleration and DLAs have made enormous progress over the last two decades and have already found applications in beam manipulation and diagnostic of conventional electron bunches generated by DC or RF guns and accelerators. We expect this will continue and actual accelerators will become available with the progress in THz source technology.

## 7     Acknowledgement

This work has been supported by the European Research Council under the European Union's Seventh Framework Programme (FP7/2007-2013) through the Synergy Grant "Frontiers in Attosecond x-ray Science: Imaging and Spectroscopy" (AXSIS) (609920) and the Cluster of Excellence "Advanced Imaging of Matter" of the Deutsche Forschungsgemeinschaft (DFG) - EXC 2056 - project ID 390715994 as well as the accelerator on a chip program (ACHIP) funded by the Gordon and Betty Moore foundation (GBMF4744).